\documentclass[nofootinbib,a4paper,aps,pra,showpacs,preprintnumbers,twocolumn]{revtex4-2}
\usepackage[T2A]{fontenc}
\usepackage{amssymb}
\usepackage{amsmath}
\usepackage{mathtools}
\usepackage{braket}
\usepackage{tikz}
\usepackage{graphicx}
\usepackage{epstopdf}
\usepackage[utf8]{inputenc}
\usepackage[russian, english]{babel}
\usepackage{url}
\usepackage{listings}
\usepackage{gensymb}
\usepackage{xcolor}
\usepackage{ulem}
\usepackage{enumitem}
\usepackage{comment}

\newlist{steps}{enumerate}{1}
\setlist[steps, 1]{label = Step \arabic*:}

\begin{document}

\title{Optimized Amplitude Amplification for Quantum State Preparation}
\author{A.A. Chernikov$^1$, K.R. Zakharova$^1$, S.S. Sysoev$^2$}
\affiliation{\center{${}^1$ Saint-Petersburg\;State\;University,}\newline ${}^2$ Leonhard\;Euler\;International\;Mathematical\;Institute}
\begin{abstract}

In this paper, we present an algorithm for preparing quantum states of the form $\sum_{i=0}^{n-1} \alpha_i |i\rangle$, where the coefficients $\alpha_i$ are specified by a quantum oracle. Our method achieves this task twice as fast as the best existing algorithm known to the authors. Such state preparation is essential for quantum algorithms that process large classical inputs, including matrix inversion and linear system solvers. The standard approach relies on amplitude amplification, a process that may require multiple, time-consuming oracle queries. Consequently, reducing the number of queries—and thereby the overall time complexity can lead to significant performance improvements in practice.

\end{abstract}

\maketitle

\hyphenation{}
	
\section{Introduction}

Let $\alpha$ be a vector in $\mathbb{R}^N$ with components $\alpha_i$, $i \in \{0, \dots N\}$. Our goal is to encode $\alpha$ into a $\log_2{N}$ qubits in the form  
$$
|a\rangle = \frac{1}{\|\alpha\|}\sum_{i=0}^{N-1} \alpha_i |i\rangle
$$

This representation is referred to as an “analog-encoded state” in \cite{MKF} and is used, for example, in \cite{HHL} to encode the vector of constant terms in a system of linear equations (SLE). The amplitudes $\alpha_i$ can be provided by an oracle function $f$ such that $f(i) = \alpha_i$. Alternatively, $\alpha$ can be given as a classical vector, in which case “oracle QRAM” is introduced as the quantum analogue of $f$. This latter setting is studied in \cite{Pr}, where an algorithm for preparing the above state is described.

The algorithm in \cite{Pr} relies on amplitude amplification \cite{AmpAmp}, which generalizes Grover’s unsorted database search \cite{Grov}. This approach achieves a quadratic speedup over the corresponding classical method (in terms of the number of oracle queries). However, it may require four oracle calls per iteration - or possibly only two - since it remains unclear in the original work whether the author found a way to avoid doubling the queries for the uncompute step. Here, we present an improved algorithm that reduces the number of oracle calls per iteration to one - without increasing the total number of iterations or compromising performance. 

The remainder of the paper is organized as follows. In Section 2, we provide a concise overview of the algorithm proposed in \cite{Pr}. Section 3 introduces our new algorithm along with a proof of its correctness. In Section 4, we explain how the uncompute step can be avoided, thereby reducing iteration complexity. Section 5 extends the reasoning in \cite{Pr} to obtain a more efficient exact solution for the state preparation problem. Finally, Section 6 offers concluding remarks and provides a link to the QisKit implementation of the proposed algorithm.

\section{The original approach}

Let $n = log_2 N$ be the number of qubits needed to store the function argument, and let $m$ be the number of qubits required to store the function values. Prakash’s approach \cite{Pr} proceeds as follows. First, an equal-weight superposition of the arguments is prepared in the first register (using, for example, the Walsh-Hadamard or Quantum Fourier transform). Next, the oracle is called to obtain the corresponding function values in the second register. A third register, consisting of a single ancilla qubit, is initialized in the state $|0\rangle$:
$$
|0\rangle^n |0\rangle^m|0\rangle \overset{H^{\otimes n}}{\rightarrow} \frac{1}{2^{n/2}}\sum_{i=0}^{N-1}|i\rangle |0\rangle^m|0\rangle \overset{U_f}{\rightarrow} 
$$
$$
\overset{U_f}{\rightarrow} \frac{1}{2^{n/2}}\sum_{i=0}^{N-1}|i\rangle |f_i\rangle|0\rangle
$$
Controlled rotations of the ancilla qubit are then performed around the y-axis, with angles proportional to the values in the second register. 
$$
\frac{1}{2^{n/2}}\sum_{i=0}^{N-1}|i\rangle |f_i\rangle|0\rangle \overset{R_y(f)}{\rightarrow} 
$$
$$
\overset{R_y(f)}{\rightarrow} \frac{1}{2^{n/2}}\sum_{i=0}^{N-1}|i\rangle |f_i\rangle(\sin{\phi_i}|0\rangle + \cos{\phi_i}|1\rangle) 
$$
$$
\phi_i = \frac{\pi f_i}{2\max_{i}|f_i|}
$$
Post-selecting $|0\rangle$ in the ancilla qubit ensures that $f_i$ is approximated by $\sin{\phi_i}$. However, because such post-selection often has a low probability of success, amplitude amplification is employed.
Let $|\omega\rangle$ be the target state, with normalization constant $A=\sqrt{N\sum_{i=0}^{N-1} \sin^2{\phi_i}}$:
$$
|\omega\rangle = \frac{1}{A}\sum_{i=0}^{N-1} \sin{\phi_i} |i\rangle |0\rangle^m |0\rangle 
$$
$$
|\omega^{\perp}\rangle = \frac{1}{\sqrt{1-A^2}}\sum_{i=0}^{N-1} \cos{\phi_i} |i\rangle |0\rangle^m |1\rangle 
$$
Define $|s\rangle$ as the superposition obtained after the oracle query and rotation, followed by uncomputing the oracle:
$$
|s\rangle = \frac{1}{2^{n/2}}\sum_{i=0}^{N-1}|i\rangle |0\rangle^m (\sin{\phi_i}|0\rangle + \cos{\phi_i}|1\rangle) 
$$
The operator that creates $|s\rangle$ applies the oracle and the rotation, then reverses the oracle:
$$
U_r = U_f^{-1} R_y U_f H^{\otimes n}
$$
A reflection $U_s$ around $|s\rangle$ can be constructed via $R_0$, which reflects the entire space around the all-zeros state. 
$$
U_s = U_r R_0 U_r^{-1}
$$
Another operator, $U_{\omega}$, acts on the ancilla qubit to invert the amplitude of $|\omega\rangle$ while leaving all orthogonal states unchanged:
$$
U_{\omega} = XZX
$$
Each iteration of amplitude amplification thus involves four calls to the oracle: 
$$
U = U_s U_{\omega}
$$
With each iteration, the angle between $|s\rangle$ and $|\omega\rangle$ is reduced by $2\theta$, with
$$
\theta = \arcsin{\sqrt{\frac{1}{N}\sum_{i=0}^{N-1} \sin^2{\phi_i}}}
$$

\section{The new approach}

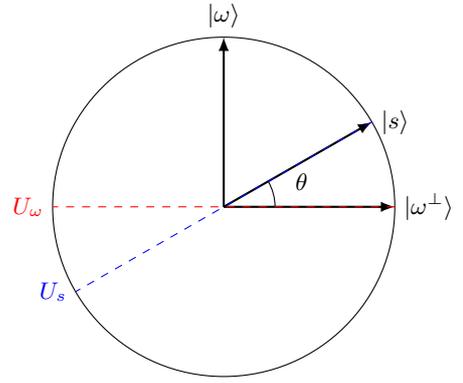
\begin{figure}
\begin{center}
\begin{tikzpicture}[scale=2.25]
\draw (0,0) circle (1);
\draw[-latex, thick] (0,0) -- (1,0) node[right] {$\ket{\omega^\perp}$};
\draw[-latex, thick] (0,0) -- (0,1) node[above] {$\ket{\omega}$};
\draw[-latex, thick] (0,0) -- (0.866,0.5) node[right] {$\ket{s}$};
\draw (0.3,0) arc (0:30:0.3) node[right=7] {$\theta$};
\draw[red, dashed] (1,0) -- (-1,0) node[left] {$U_\omega$};
\draw[blue, dashed] (0.866,0.5) -- (-0.866,-0.5) node[left] {$U_s$};
\end{tikzpicture}        
\end{center}
    \caption{Vectors $|\omega\rangle$, $|\omega^{\perp}\rangle$ and $|s\rangle$ lie in the same plane, where the action of $U_\omega$ and $U_s$ represent reflections around the corresponding axes.}
    \label{fig:grover}
\end{figure}

As described in the previous section, amplitude amplification can be visualized as a rotation in the two-dimensional subspace spanned by $|\omega\rangle$ and $|\omega^{\perp}\rangle$. The operator $U_{\omega}$ reflects the state about $|\omega^{\perp}\rangle$, whereas $U_s$ reflects it about $|s\rangle$ (see Fig.~\ref{fig:grover}).

In our approach, we define the vectors as follows:
$$
|\omega\rangle = \frac{1}{A}\sum_{i=0}^{N-1} \sin{\phi_i} |i\rangle |0\rangle^m (e^{i\phi_i}|0\rangle - e^{-i\phi_i}|1\rangle) 
$$
$$
|\omega^{\perp}\rangle = \frac{1}{\sqrt{1-A^2}}\sum_{i=0}^{N-1} \cos{\phi_i} |i\rangle |0\rangle^m (e^{i\phi_i}|0\rangle + e^{-i\phi_i}|1\rangle) 
$$
$$
|s\rangle = \frac{1}{2^{n/2}}\sum_{i=0}^{N-1}|i\rangle |0\rangle^m |+\rangle
$$
where $A$ is a normalization constant. 

The reflection about $|\omega^{\perp}\rangle$ is performed using two oracle queries via the operator
$$
U_\omega = U_f^{-1}XR_z(2f_i) U_f
$$
Its action on $|\omega\rangle$ proceeds as follows:
$$
|\omega\rangle = \frac{1}{A}\sum_{i=0}^{N-1} \sin{\phi_i} |i\rangle |0\rangle^m (e^{i\phi_i}|0\rangle - e^{-i\phi_i}|1\rangle) \overset{U_f}{\rightarrow}
$$
$$
\rightarrow \frac{1}{A}\sum_{i=0}^{N-1} \sin{\phi_i} |i\rangle |f_i\rangle (e^{i\phi_i}|0\rangle - e^{-i\phi_i}|1\rangle) \overset{R_z(2f_i)}{\rightarrow}
$$
$$
\rightarrow \frac{1}{A}\sum_{i=0}^{N-1} \sin{\phi_i} |i\rangle |f_i\rangle (e^{-i\phi_i}|0\rangle - e^{i\phi_i}|1\rangle) \overset{X}{\rightarrow}
$$
$$
\rightarrow \frac{1}{A}\sum_{i=0}^{N-1} \sin{\phi_i} |i\rangle |f_i\rangle (e^{-i\phi_i}|1\rangle - e^{i\phi_i}|0\rangle) \overset{U_f^{-1}}{\rightarrow}
$$
$$
\rightarrow \frac{1}{A}\sum_{i=0}^{N-1} \sin{\phi_i} |i\rangle |0\rangle^m (e^{-i\phi_i}|1\rangle - e^{i\phi_i}|0\rangle) = -|\omega\rangle
$$
One can verify that $U_\omega$ acts as the identity on $|\omega^{\perp}\rangle$.
Meanwhile, the reflection about $|s\rangle$ requires no oracle queries and is defined by
$$
U_s = H^{\otimes n}R_0H^{\otimes n}
$$
Thus, a single amplitude amplification iteration is $U=U_s U_\omega$. Each iteration reduces the angle between the system’s current state and $|\omega\rangle$ by $2\theta$, where
$$
N\sin^2{\theta} = \sum_{i=0}^{N-1} \sin^2{\phi_i}
$$
and
$$
|s\rangle = \sin{\theta} |\omega\rangle + \cos{\theta} |\omega^{\perp}\rangle 
$$
Since $\theta$ remains the same as in the previous section, the total number of required iterations does not change. However, this new approach reduces the number of oracle queries per iteration by two.

\section{Avoiding the Uncompute}

In the previous section, each call to the oracle was paired with its inverse in order to clear the function value register, effectively doubling the number of oracle queries. However, this overhead can be avoided. In this section, we outline how to achieve this.

Define the operator $U_f$ by
$$
U_f |x\rangle |y\rangle = |x\rangle |y + f(x)\rangle
$$
Implementing $U_f$ requires a quantum addition algorithm alongside an oracle query, and it can be realized with $O(m^2)$ elementary gates. Our aim is to use the Quantum Fourier Transform (QFT) to build the operator $U_\omega$ with just one invocation of $U_f$. Let $M = 2^m$. Then:
$$
|i\rangle|1\rangle^{m} \overset{QFT_m}{\rightarrow} |i\rangle \frac{1}{\sqrt{M}} \sum\limits_{k=0}^{M-1} e^{-i\frac{2\pi k}{M}} |k\rangle \overset{U_f}{\rightarrow} 
$$
$$
\rightarrow |i\rangle \frac{1}{\sqrt{M}} \sum\limits_{k=0}^{M-1} e^{-i\frac{2\pi k}{M}} |k + f(i)\rangle =
$$
$$
= |i\rangle \frac{1}{\sqrt{M}} \sum\limits_{k=0}^{M-1} e^{-i\frac{2\pi}{M} (k - f(i))} |k\rangle = $$
$$
=e^{i\frac{2\pi}{M} f(i)} |i\rangle \frac{1}{\sqrt{M}} \sum\limits_{k=0}^{M-1} e^{-i\frac{2\pi k}{M}} |k\rangle \overset{QFT_m^{*}}{\rightarrow} 
$$
$$
\rightarrow e^{i\frac{2\pi}{M} f(i)}|i\rangle|1\rangle^{m}
$$
To obtain a phase of $e^{-i\frac{2\pi}{M} f(i)}$, one may replace $QFT$ with $QFT^{*}$ in the above sequence. Whether to use the direct or inverse QFT can be guided by an ancilla qubit, which ensures the correct implementation of the previously defined operator $U_\omega$.

\section{The exact solution}

Prakash \cite{Pr} presents an approach to obtain an exact solution to the problem, meaning that the probability of $|\omega\rangle$ becomes 1 after a set number of amplitude amplification iterations. In that approach, an angle $\overline{\theta}$ is introduced such that $\frac{\pi}{2\overline{\theta}}$ is an integer. If $p=\sin^2{\theta}$ is the original probability of $|\omega\rangle$ after the oracle query, then $\overline{p}=\sin^2{\overline{\theta}}$ is the probability with the modified oracle. Prakash’s method requires adding an ancilla qubit initialized in the superposition
$$
\sqrt{\frac{\overline{p}}{p}} |0\rangle + \sqrt{1-\frac{\overline{p}}{p}} |1\rangle
$$
and modifies $U_\omega$ to reflect about the state $|00\rangle$ (i.e., both ancillas in the zero state).

In contrast, our new approach does not require any additional ancillas. The angle $\theta$ is adjusted directly within the oracle query:
$$
U_\omega = U_f^{-1}XR_z(2f_i \frac{\overline{\theta}}{\theta} ) U_f
$$
which provides a simpler and more direct way to modify the rotation angle.

\section{Conclusion}

In this paper, we introduced an approach that achieves analog-encoded state representation at twice the speed of the best modern algorithm presented in \cite{Pr}, using oracle-query complexity as the performance metric. Additionally, our method offers a more streamlined, memory-efficient means of obtaining an exact solution. We believe these advantages make the proposed technique significantly more suitable for practical implementations.

The algorithm implementation in QisKit can be found here: \url{https://colab.research.google.com/drive/1JkxFG7Mr2Ukz5NKlVAPGA9WX-oKwR-AD?usp=sharing}

\section*{Acknowledgments}
The work is supported by Ministry of Science and Higher Education of the Russian Federation, agreement № 075–15–2022–287.

\end{document}